# Compact Heterogeneous Integration of Scalable Array with Image Selection Architecture for RFIC Circuitry to Miniaturized Power Delivery Network on the Package for the Next Generation of 6G and Beyond

Najme Ebrahimi, *Member, IEEE*

***Abstract*—Next generation communication and sensing require enabling technologies for miniaturized and efficient heterogeneous systems while integrating technologies ranging from silicon to compound semiconductors and from photonic chips to micro-sensors. To this end, high frequency and mm-wave (MMW) lossy parasitics and delay between modules need to be significantly reduced to minimize area, loss and thermal heating of inter-chip wiring and power delivery networks. In this work, we propose novel approaches to achieve an efficient wideband MMW array integrations. The proposed techniques are built upon the following: 1) fixed antenna package buildup for every element with differential excitation on two half sides of array to reduce the fabrication cost and the IC-to-antenna routing loss; 2) miniaturized aperture coupled local oscillator (LO) and intermediate frequency (IF) power delivery feed distribution to minimize the packaging stacked layers and their loss. The proposed 16-element antenna array is integrated which 4 dies in 2x2 configurations implemented in a 90-nm SiGe BiCMOS process using compact Weaver image-selection architecture (WISA). The proposed miniaturized and efficient architecture from circuit and chip level to package level results in 1.5 GHz modulation bandwidth for 64 QAM (9 Gb/s) and 2 GHz for 16 QAM with only ±2 dB EVM variation over the 20% FBW (71-86 GHz). The system produces 30-dBm EIRP with enhanced efficiency of 25% EIRP/PDC over the bandwidth.**

## I. INTRODUCTION

FUTURE heterogeneous mm-wave (MMW) systems, 6G and beyond, for next generation of communication and sensing, require different technologies and components including memory, microprocessors, Micro-electromechanical (MEMS) and other passive and optical sensors into a higher-level System-in-Package (SiP). This heterogeneous system should contain MMW and THz scalable antenna array to enable transmitting or receiving data to the environment, Fig. 1. The future MMW and THz scalable array also necessitates power efficient and compact 2-D/2.5D to 3-D integration of various technologies ranging from silicon to compound semiconductors (*III-V* materials). The miniaturized and efficient integration of such system requires novel architectures and innovations from transistor level up to package and system level. One of the major challenges for future heterogeneous integration is interconnects and power delivery network architectures to enable efficient power distribution with miniaturized layout and low-phase and amplitude mismatch over wide bandwidth. To this end, MMW and high frequency lossy parasitics and delay between modules in the scalable array also need to be significantly reduced to minimize area, loss and thermal heating of inter-chip wiring and power delivery networks. State-of-the-art (SoA) approaches use low-loss substrates technologies such as low-temperature co-fired ceramics (LTCC) that can support multi-layers for the dense routings and either wire-bond [1-6], flip-chip [7-9], embedded wafer level ball grid array (BGA) or fan-out wafer level packaging for interconnection between modules to package, [10-12]. They employ multiple stacked wiring layers that induce large loss and delay due to the parasitics of long vias and interconnects. In particular, for MMW array above 60 GHz, the SoA reported a fill factor less than 70% with a DC power consumption per elements varying from minimum 60 mW for a 60 GHz band [13-15] to 275 mW for 80-100 GHz band [16-30], resulting in the efficiency EIRP/PDC ranging from 15% to 10%, respectively.

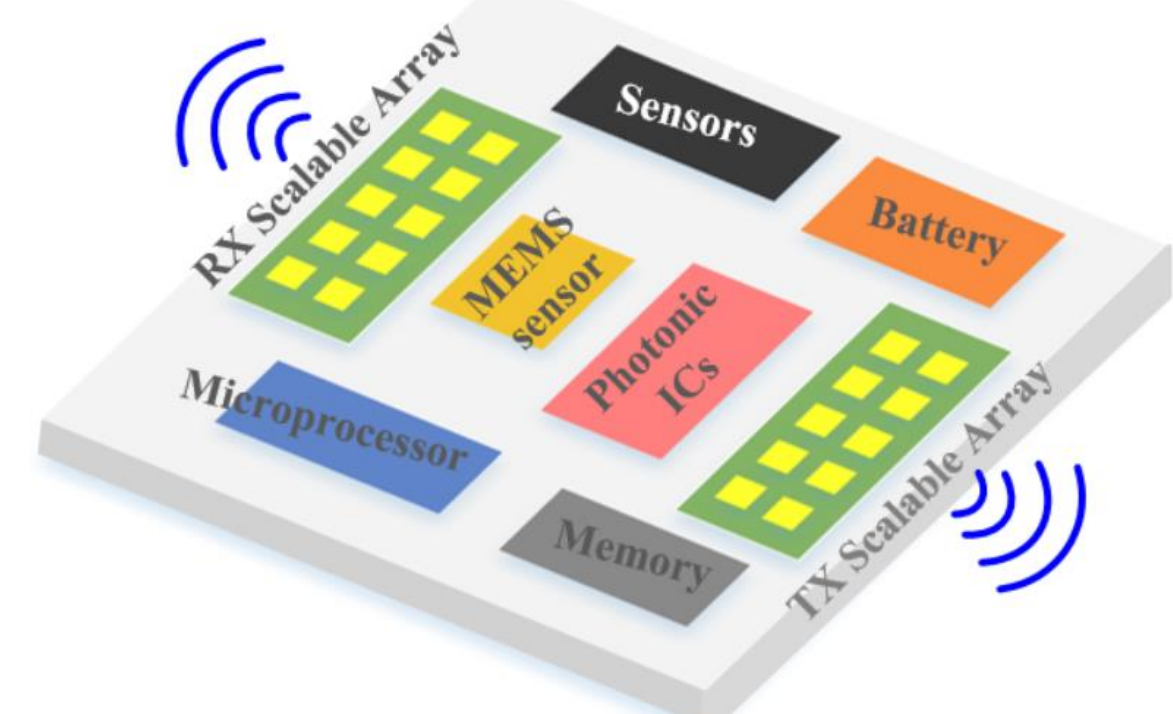

Fig 1. Future heterogeneous integration containing different technologies and components including memory, microprocessors, MEMS and other passive and optical sensors integrated with MMW and THz scalable antenna array to enable transmitting or receiving data to the environment.

In this work, novel approaches for scalable MMW power delivery network are proposed to achieve an efficient wideband MMW array, attaining 20% fractional bandwidth (FBW) from 71-86 GHz, *E*-band, with EIRP/PDC of 25% and with the minimum performance variation over bandwidth. The proposed techniques are built upon the following: 1) fixed antenna element for every element with differential excitation on two half sides of array to reduce the fabrication cost and the IC-to-antenna routing loss; 2) miniaturized aperture coupled local oscillator (LO) and intermediate frequency (IF) power delivery feed distribution to minimize the packaging stacked layers to only 4-layers and their loss. Aperture coupling technique has been traditionally used for wideband antenna feed network. However, in this work, for the first time, aperture coupling technique is employed for wideband and low- phase and amplitude mismatch power feed distribution with simultaneous impedance matching and impedance transformation between *M*-dies in an *N*-element array , where $(N = m \times M)$ and $m$ is number of elements per die.

Section II describes the fixed antenna unit concept for every element to reduce the interconnect loss from RFIC package to antenna package. The proposed area and power efficient LO/IF power delivery network is discussed with scalable impedance matching feature in Sec. III. Section IV discusses the analytical expression to employ the aperture coupled technique for efficient power distribution with simultaneous impedance transformation at every intermediate nodes of an *N*-element array. The measurement results for the 16-elements antenna array integrated with four die in 2x2 configuration, fabricated in 90nm BiCMOS, are presented in Sec. V. Section VI concludes the paper

## II. Fixed Antenna Package Buildup with Differential Excitation on Two Half-Sides of Array

In order to reduce the interconnect loss from RFIC package to antenna package, the interconnects need to be made short. According to the SoAs at three different bands, 60 GHz, [10-15], and 80-100 GHz, [16-30], and 28 GHz [31-36], the routings between IC to antenna package typically introduce 2-4 dB loss per element with ±1 dB variation between elements. In general, for every 2 dB RF routing loss, the fill factor of array, equals to $N_{eff}/N_{org}=(10^{-Ploss/20})$ %, is reduced by 20% resulting in effective array element, $N_{eff}$, of 204 in an array with original elements, $N_{org}$, of 256 for base station applications or $N_{eff}$=13 for mobile users array with $N_{org}$=16. On top of the RF routings, the IF and LO power delivery networks introduce an average loss of 15 dB, [13], [16], [34-36] and 2° phase mismatch using a daisy chain master-slave network, Fig. 2(a). Therefore, to compensate the loss or phase/amp mismatch in the packaging routings, additional amplifier gain stages on each IC/die is required. For every 2 dB gain compensation at 20 GHz frequency for LO and 100 GHz at RF, each amplifier typically consumes 20mW and 40mW, respectively, [37-41], introducing extensive heating to the scalable array system. In this work, a fixed antenna package for all dies are employed, where each antenna is excited through a patch and microstrip lines with differential excitation for two half sides of arrays for symmetric and short layout, as shown in Fig. 1(b). This will reduce the antenna package routing loss as well as the fabrication and characterization costs as all the elements have fixed antenna feed lines. To compensate the differential excitation on two half array sides in the fixed antenna package, additional differential phase shift is required in the package. Therefore, a novel compact differential power delivery network is proposed to be employed on IF or LO path. The proposed feed network also generates differential signal based on the aperture coupling with λ/4 open stub, [6], (see Sec. III and IV for more details). This proposed layout reduces the number of interface layers to only 4-layer for other signals routings such as I/O, digital and power supplies. The simulated realized gain for 16-element antenna array for both conventional and proposed approaches are illustrated in Fig 3 (a) and (b), for 86 GHz and 71 GHz, respectively. The simulated antenna gain for the proposed architecture varies from 15 dB gain to 17.3 dB from 71 GHz to 86 GHz (below 2dB variation). However, the proposed architecture enhances the insertion loss of IC to antenna routings 1.5 dB at 71 GHz and 2.5 dB at 86 GHz. Note that every 2 dB enhancement in insertion loss will save the fill factor by 20% and $P_{DC}$ of PA by 40 mW for 100 GHz array. In addition to saving the production cost, the proposed package layout reduces the process of hardware assembly, screening, and testing for characterization as all the routings in the antenna package is the same for each element with only opposite rotation.

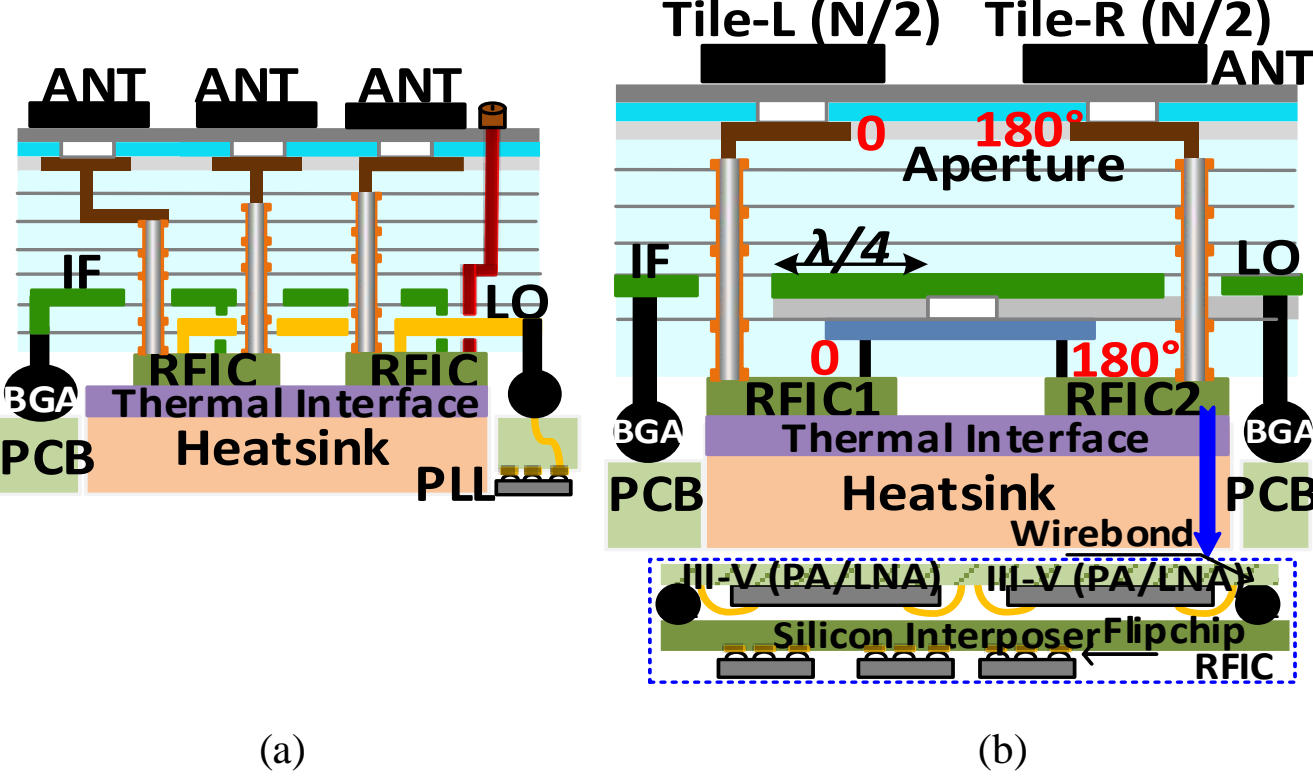

(a) (b)

Fig 2. Heterogeneous antenna-IC 3D integrations, a) conventional layout with different excitation for antenna feed line. b) proposed package layout with fixed antenna package, novel LO/IF distribution,

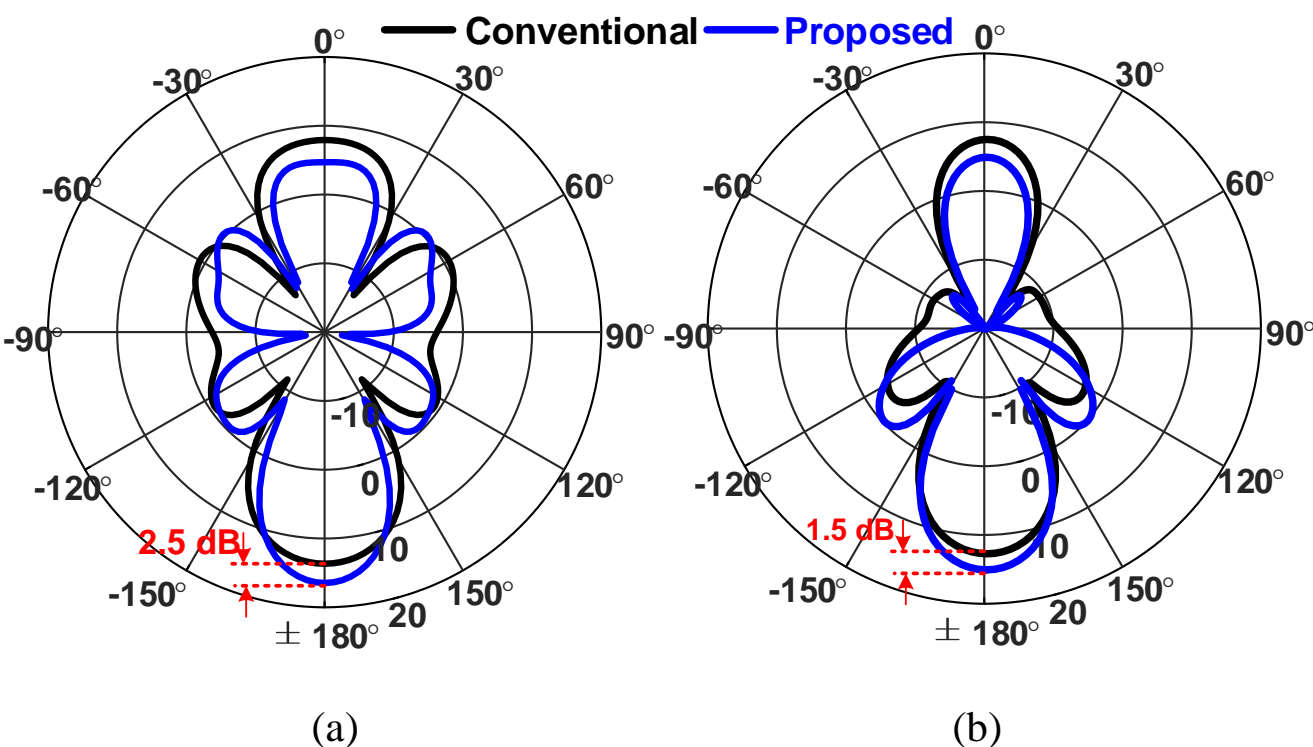

(a) (b)

Fig 3. Simulated 16-element antenna array gain for conventional and proposed layout in (yz-plane), a) 86 GHz, 2.5 dB gain enhancement, b) 71 GHz, 1.5 dB gain enhancement.

## III. High Frequency Power Delivery Network For MMW Array

In addition to the requirement of the short and symmetric routings for RF path (IC-to-antenna), the LO signal generated by a PLL and IF signals need to be distributed to each individual dies of the array through a low-loss, low amp/phase mismatch and small foot print power delivery network. The two main schemes for power delivery networks are daisy chains and H-tree with trade-offs between the phase mismatch/phase noise and scalability. The H-tree has demonstrated better scalability and phase/amp matching between dies [13, 16, 42]. However, in order to compensate for the routing loss for a large number of arrays, accurate impedance matching over LO and IF frequency bandwidth is required through the distribution network. Without a matched environment and matched transmission line, the distribution signals will have standing waves at various points and therefore, the received power and LO amplitudes at each dies depend on the load impedance and physical layout of the entire distribution network. The two typical H-tree power delivery networks are simple T-junction and Wilkinson as shown in Fig. 4(b) and (c), respectively, for an $N$-element array containing $M$-dies supporting $m$-element each, ($N=m\times M$). The simple T-junction network, shown in Fig. 4(b) limits the number of dies, $M$, due to the impedance matching constraints as increasing the number of dies creates smaller impedance at each junction and finally at PLL input ($Z_0/M$). Therefore, Wilkinson power delivery network, Fig. 4(c), is a desired option for large element arrays as it provides 50Ω impedance matching at every junction with enough isolation. However, it has large foot print due to $\lambda/4$ arms and their associated loss, while having the requirement of adding impedance layers to the package to create $2Z_0$ resistance between the two dividing ports. Furthermore, in our proposed miniaturized and low loss packaging scheme presented in Sec. II, a180° differential phase shift to the system in IF or LO power delivery path is needed. On-chip balun or conventional rat-race couplers are poor candidates for compact differential feed network to meet spacing requirements between antennas. The novel power distribution network based on aperture coupling technique, previously proposed by authors in [6], are used for large scalable array integrations with inherent differential phase shift. Aperture coupling is also inherently wideband (20% FBW) and low-loss by removing lossy vias [43,44]. The proposed feed network uses three layers as illustrated in Fig. 4 (d). The bottom microstrip feed line is open stub and the aperture is approximately $\lambda/4$ from the end of the line to ensure short circuit at the aperture for maximum current injection and magnetic coupling. It can also employ the top layer as a H-tree combiner/devider while integrated to the impedance matching transformation. To calculate and to compare the area and loss of these networks, the estimated routing length $l_{r\text{-}total}$ should be calculated. For $M$-die supporting $m$-element creating $N$-element array ($N= M\times m$) with given antenna element spacing of $d$, the approximated center-to-center spacing of each die, $D$, is $m\times d$. As shown in Fig. 4(a), the routing length, $l_{r\text{-}total}$ is the summation of $d_i$'s each incrementing by $D$ as number of dies multiplied by two. Therefore,

$$d_i = (m \times d)\lfloor (i+1)/2 \rfloor, \qquad \text{for } i = 0 \text{ to } \log_2(M), \tag{1}$$

and the total routing is approximated by:

$$l_{r-total} = \sum_{i=0}^{log_2^M} \left\lfloor \frac{M}{2^i} \right\rfloor \cdot D_i \approx \frac{4}{3} N \times d \tag{2}$$

This can be multiplied by the loss of a transmission line in dB/mm, $\alpha_0$, to estimate the total loss and be multiplied by width, $W$, of T-line to estimate the foot-print. In Fig. 4 (e), these parameters are calculated and are compared between different techniques and the proposed technique. Note that the associated loss of dividing junction, 3 dB, and its foot print proportional to $\lambda_w/4$ for Wilkinson and $\lambda_g/4$ *for* open stub, should be also considered and be multiplied by the dividing junction number. The $\lambda_w$ and $\lambda_g$ are the wavelength of the transmission line in the Wilkinson and proposed approach, respectively. The number of dividing junctions for Wilkinson is

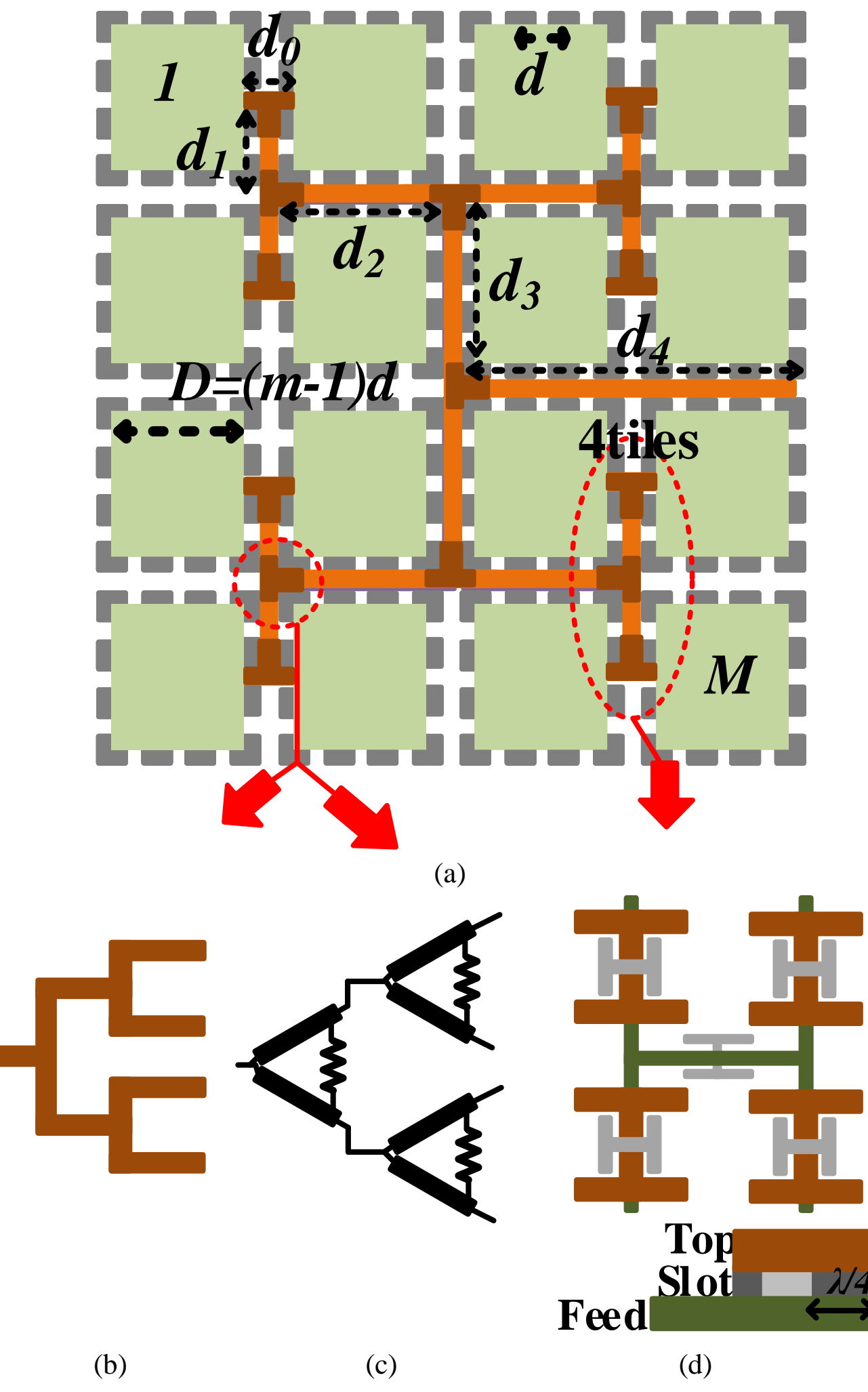

| | H-tree T-junction | H-tree Wilkinson | Proposed |
|---|---|---|---|
| Foot-print (Fig. 5 (a) | $\frac{4}{3}dNW$ | $\left(\frac{4}{3}dN + \frac{\lambda_W}{4}M\right)W$ | $\boldsymbol{\left(\frac{4}{3}dN + \frac{\lambda_g}{4}\frac{M}{3}\right)W}$ |
| Loss (Fig. 5 (b) | $\frac{4}{3}dN\alpha_0$ | $\frac{4}{3}dN\alpha_0 + 3 + 10log_{10}(M)$ | $\boldsymbol{\frac{4}{3}dN\alpha_0 + 3 + 10log_{10}(M/3)}$ |
| Scalability | Low[1] | Medium[2] | **Highest[3]** |
| Phase Shift | 0 | 0 | **0/180°** |

1.Create small impedance as $N$ increases, 2. Require resistance $2Z_0$, 3. 50Ω match at each junction

(e)

Fig 4. a) LO or IF power delivery network layout in $N=m\times M$ element array with M-dies/IC covering $m$-element antenna, with three different techniques; b) T-junction, c) Wilkinson and d) proposed differential aperture coupling technique, e) Comparison Between Different Power Delivery Network

$$\sum_{i=1}^{log_2^M} \left\lfloor \frac{M}{2^i} \right\rfloor \approx M \tag{3}$$

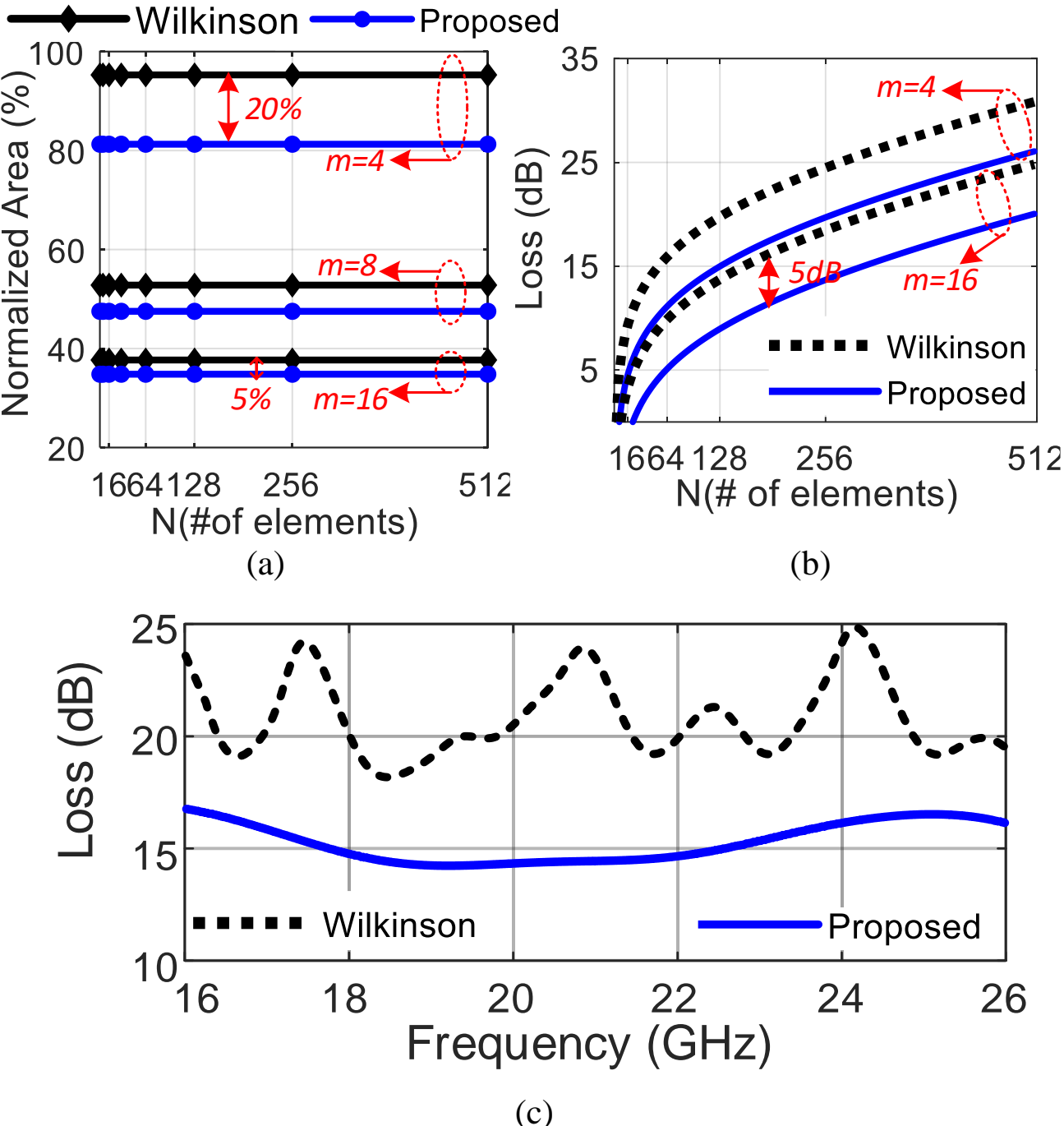

(a) (b)

(c)

Fig 5. a) Simulated results and comparison between Wilkinson based power delivery network and proposed aperture coupled technique, a) Normalized area vs $N$ for different $m$=4, 8, & 16, b) Loss vs $N$, c) Loss vs frequency.

Which is reduced to (4) for the proposed aperture technique as it can combine/divide four dies using the H-shape divider on the top metal line.

$$\sum_{i=1}^{log_2^M} \left\lfloor \frac{M}{4^i} \right\rfloor \approx M/3 \tag{4}$$

The calculated normalized area (power delivery network area/ total antenna elements area) and loss versus the number of elements, $N$, are also plotted in Fig. 5(a) and Fig. 5(b), respectively. As it is shown in Fig. 5(a), the proposed technique saves the area by 20% for a 2x2 chip configuration ($m=4$), as suggested in [34,45], but as the number of channels per die is increased to a 16-channel, it saves the normalized area by only 5% compared to Wilkinson divider. However, the proposed approach reduces the loss of distribution as $m$ increases, 3dB improvement by doubling the $m$, and saves the general distribution loss by 5 dB compared to Wilkinson for the 20 GHz LO band as shown in Fig. 5(b), equivalent to saving up to 50 mW for power distribution. The two power delivery networks are also simulated for $M=16$ elements in Advanced Design System (ADS) and the resulting plots are shown in Fig. 5(c) for characterizing the performance over frequency and bandwidth. The proposed technique provides 5 dB enhancement in total loss for a 16 elements with minimum variation of ±0.5 dB over 20% FBW.

## IV. Aperture Coupling Technique for Scalable Impedance Matching and Transformation for LO/IF Power delivery Network

Aperture coupled feed line contains three layers, the radiating top metal layer on top, the feed network with an open stub on the bottom layer and the small aperture slot located under the radiating elements that allows coupling between top layers and the feed line, Fig. 6 (a). The bottom microstrip feed line is open stub and the aperture is approximately $\lambda/4$ from the end of the line to ensure short circuit at the aperture for maximum current injection and magnetic coupling. We can also achieve differential phase shift from the structure due to feature of the travelling wave in opposite direction. According to the $E$-field distribution through the aperture shown in Fig. 6 (a), the signal distribution across the aperture and its top substrate will have the same amplitude but 180° out of phase, generating differential signals. This phase shift is required for our proposed architecture for scalable array with fixed antenna element (Sec. II). The magnetic coupling is strongly dependent to the size and geometry of the slot, $L_S$, $W_S$. In addition, the impedance transformation and matching condition can be controlled by the size, position and shape of the aperture as well as the open-ended stub length and the width of top radiation layer, $W_T$, Fig. 7 (a). The electrical circuit model of the aperture coupled circuitry is shown in Fig. 7 (b) with the input impedance of the top layer, $Z_L$ parallel with aperture parasitics, inductance, $L_c$ and capacitance, $C_c$, that transform to the input impedance, $Z_s$, with impedance transformation turn ratio can be expressed as

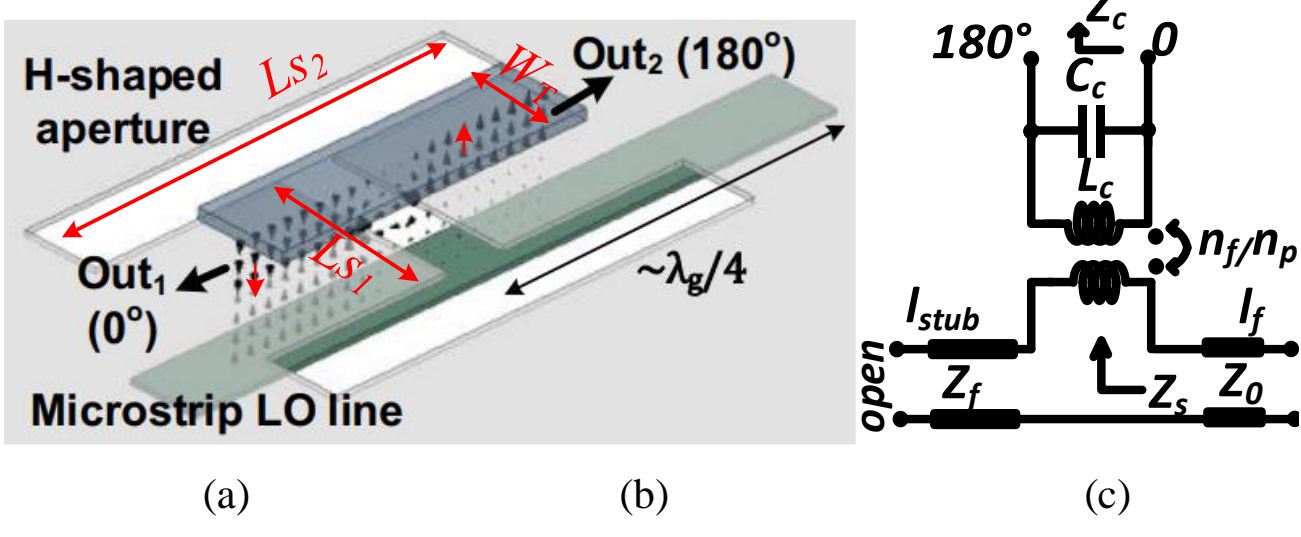

(a) (b) (c)

Fig 6. a) Proposed differential aperture coupling technique with $E$- field distribution for power delivery network, b) electrical circuit model and impedance transformation for aperture coupled.

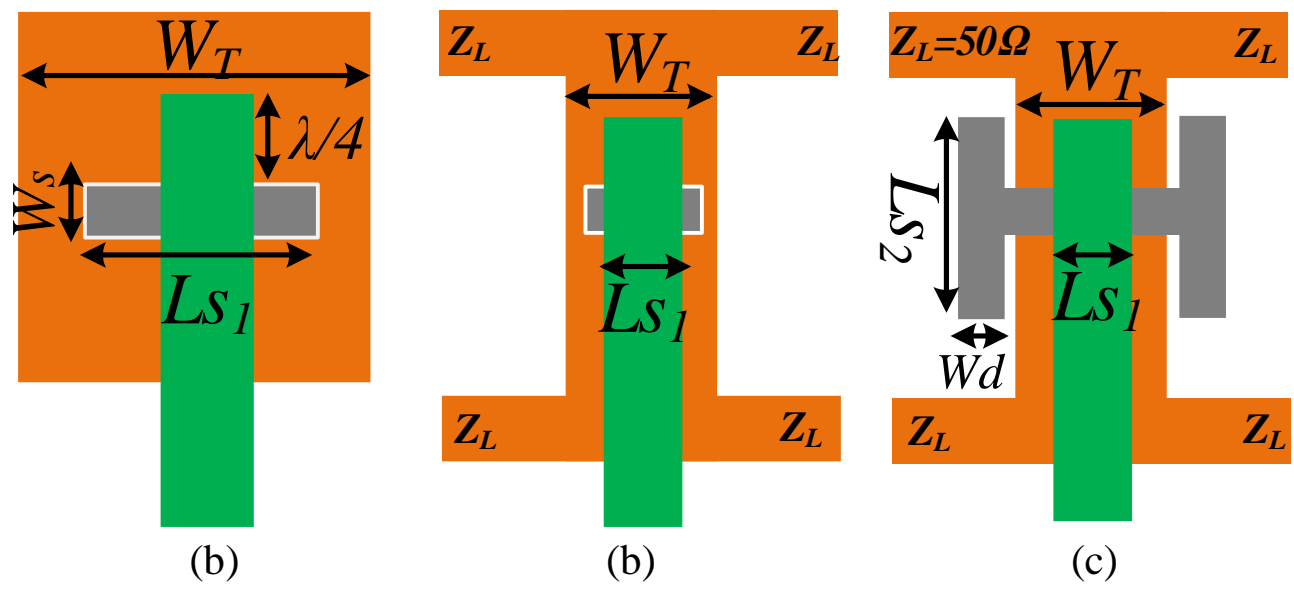

(b) (b) (c)

Fig 7. Aperture coupled different layout, a) wideband patch antenna ($L_{S1}/W_T$<<1), b) rectangular slot for H-tree power distribution ($L_{S1}/W_T$≥1), c) proposed H-shaped slot with H-tree power distribution for top layer.

$$Z_s \propto \left(\frac{n_f}{n_p}\right)^2 Z_L \tag{5}$$

where $n_f$ is the impedance turn ratio between to top layer and aperture slot, $n_p$ is the turn ratio between bottom feed line and slot, shown in Fig 6 model. The matched input impedance, $Z_{in}$, depends on the relative inductive coupling ratio between the two coupled substrate lines and aperture impedance as:

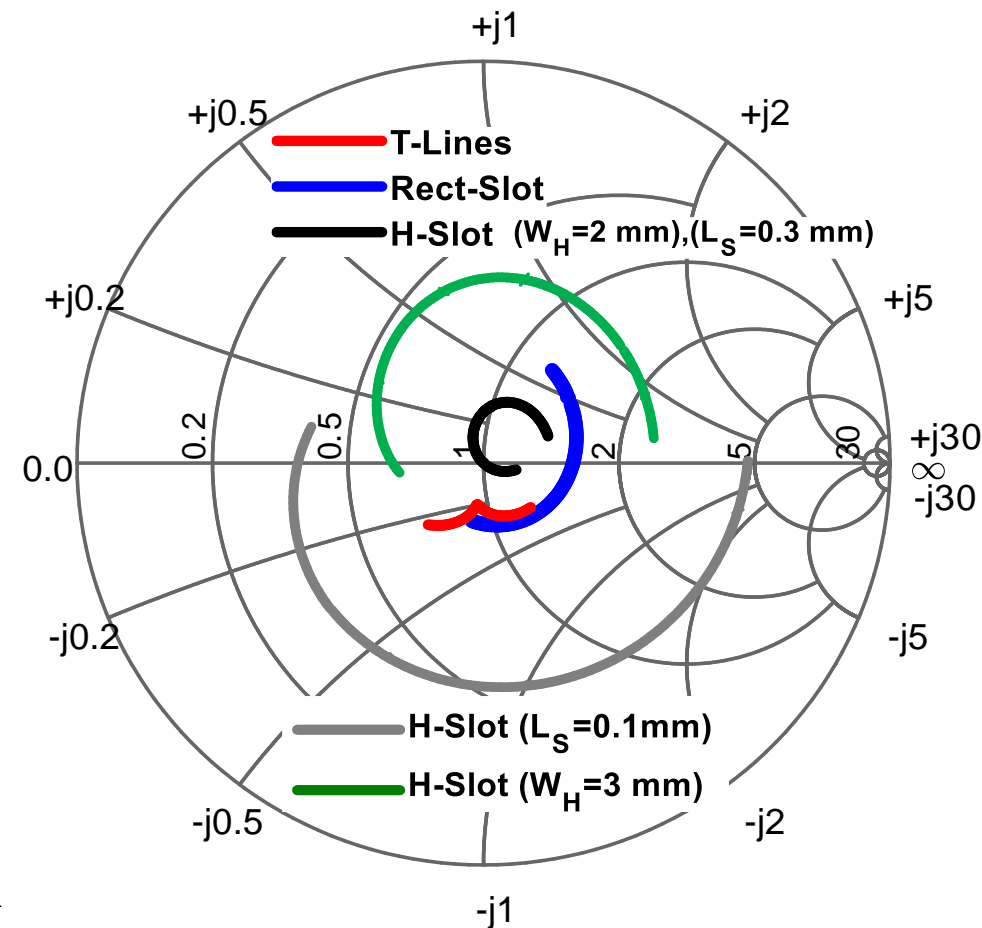

**Fig. 8.** Impedance transformation and locus points for different layout and geometry of aperture coupled circuitry.

$$Z_{in} = Z_s - jZ_0 \cot\left(\beta_f L_{stub}\right) \qquad (6)$$

Where $L_{stub}$ is the open stub length which is $\lambda/4$, with propagation constant of $\beta_f$ and transmission line characteristic impedance of $Z_0$, creating the reactive component of impedance. The $n_f$ and $n_p$ are related to the geometry of the aperture slot, $L_S$, $W_S$, the top layer width, $W_T$, and its substrate thickness, $h_i$ as [43,44]: .

$$n_p \propto \frac{L_S}{2W_T} \qquad (7.a)$$

$$n_f \propto 1 - \exp\left(-\frac{L_S}{4h_T}\right) \qquad (7.b)$$

The impedance ratio of $n_f$ is usually near 1 when the maximum coupling is achieved. Therefore, the ratio of slot length and top-layer width, $L_S/W_T$, mainly sets the impedance transformation ratio. To ensure this accurate impedance transformation, a maximum coupling, $C_c$, should be also attained ($S_{21}=1$) by tuning and optimizing aperture slot as [43,44]:

$$C_c \propto \frac{\sqrt{\varepsilon_r}}{\sqrt{\left(\frac{\pi}{W_S}\right)^2+\left(\frac{\pi}{L_S}\right)^2}} \qquad (8)$$

Where $\varepsilon_r$ is the transmission line permittivity and $W_s$ is the width of the aperture. Equations (7) and (8) indicates that there is a trade-off between the coupling and the impedance matching, which the larger size of aperture slot length, $L_s$, enhances the coupling, while reducing the impedance transformation ratio.

Conventionally, making top layer radiation patch larger, $W_T$, than the aperture slot length, $L_S$, ($L_S/W_T \ll 1$) will mitigate the coupling and impedance transformation trade-off, Fig. 7 (a). However, employing aperture coupling technique for power distribution network suffers from the comparable size of top layer and the slot length, *i.e* (($L_S/W_T \geq 1$), Fig. 7 (b). This is mainly because the top layer width, $W_T$, and size are designated by the H-tree power combiner/divider to provide impedance matching between *n*-dies of *M*-array. In order to mitigate this trade-off and add more degree of freedom to the design an H-shape aperture slot has been employed, Fig. 7 (c), for simultaneous power combining/splitting between *n*-dies and impedance transformation ratio of *n* ($Z_c=Z_0/n$).

By employing an H-shaped aperture slot, additional arms in the slot, $L_{S2}$ in Fig. 6(c), will be added to the design to benefit the enhancement of coupling by increasing the aperture size while making the top layer with smallest area, which is applicable for our case with (($L_{S1}/W_T \leq 1$). For the $L_{S1}/W_T$ of 1, equal top layer and slot horizontal length sizes, the $n_p$ is 0.5 based on (5) and with a typical $n_f$ of 1, the expected transformation ratio is 4. Therefore, the top layer H-tree power divider is a 4-way power delivery network as the proposed layout dimension can inherently support *n=4* die for impedance transformation of *4*. To assure the accurate impedance matching over the bandwidth the coupling should be also maximized too. This will happen by tuning and optimizing the *H*-arm of the aperture slot, $L_{S2}$. The simulated locus points for different aperture size are shown in the smith chart in Fig. 8. The aperture length $L_{S1}$, $L_{S2}$, are the main factors tuning the resonance frequency and affecting the impedance matching conditions. Increasing the $L_{S1}$ decrease the resonance frequency due to the increasing the capacitance parasitics of the main aperture slot, while increasing the H-arm length, $L_{S2}$, increase the coupling ratio, $C_c$, to compensate the impedance match and coupling. The optimum values of $L_{S1}$ and $L_{S2}$ of 0.3mm and 2mm are chosen to provide optimum coupling and impedance matching over the bandwidth. **Fig. 8** also compares the proposed power delivery network with the conventional rectangular aperture slot and H-tree with T-junction distributer. It is illustrated that the proposed approach can achieve the best impedance matching ratio over the bandwidth. The input return loss, $S_{11}$, of the three techniques are also compared in **Fig. 9 (a)**, depicting the proposed technique can achieve better than 20 dB return loss with tunability of resonance frequency from 20GHz to 26GHz, 5G frequency bands. This illustrates that, the proposed approach can be also employed for radio frequency (RF) power signal distribution for RF based phase shifting for future 5G. The amplitude and phase mismatch of the three techniques are also compared and shown in **Fig. 9 (b),** expressing minimum variation over bandwidth, with less than ±0.25° phase mismatch for the proposed layout.

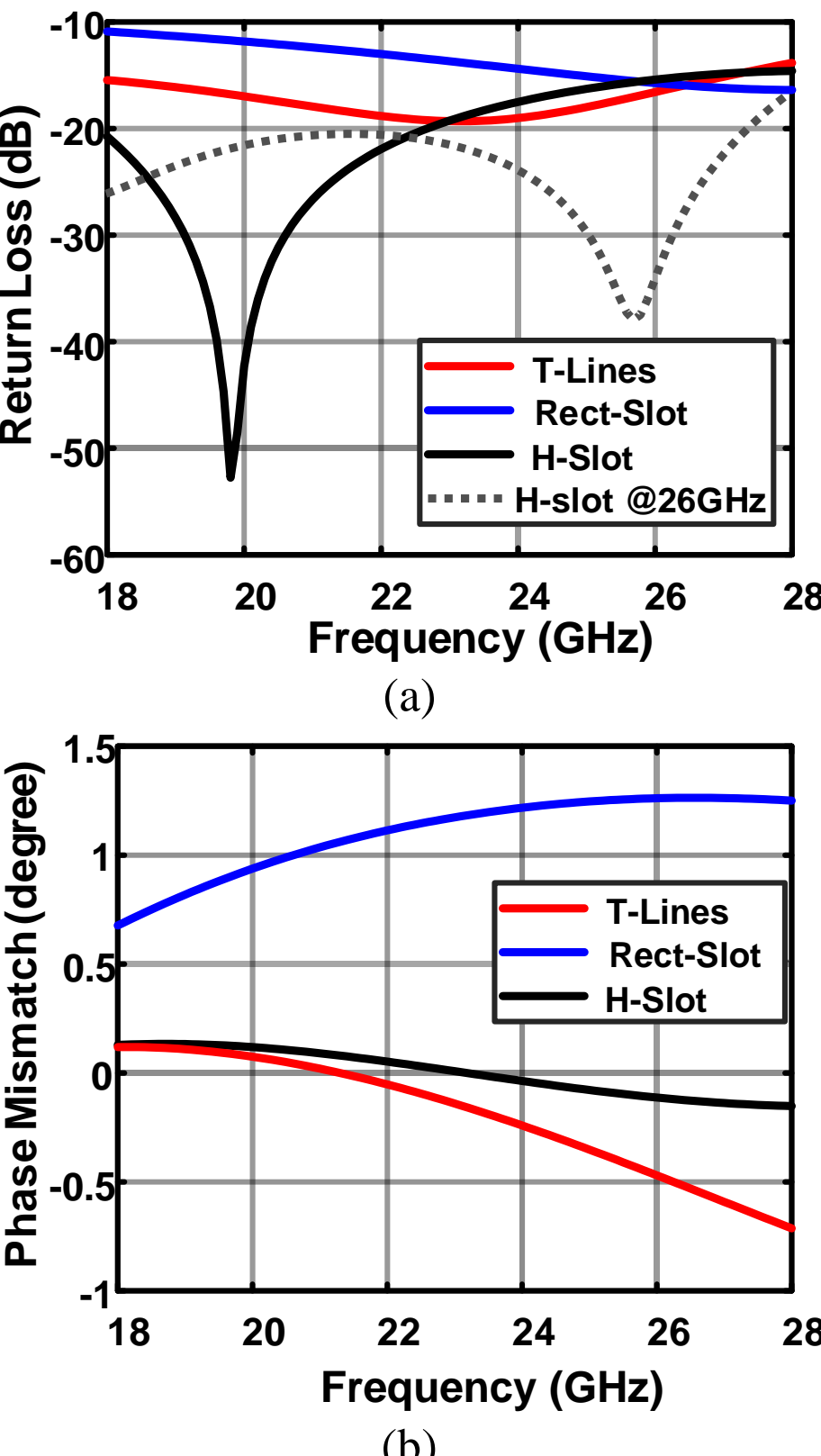

Fig. 9. S-parameter simulation and comparison between proposed H-shaped aperture and other power distribution network and layout (T-junction and aperture with rectangular slot), a) input return loss simulation versus frequency b) phase error versus frequency.

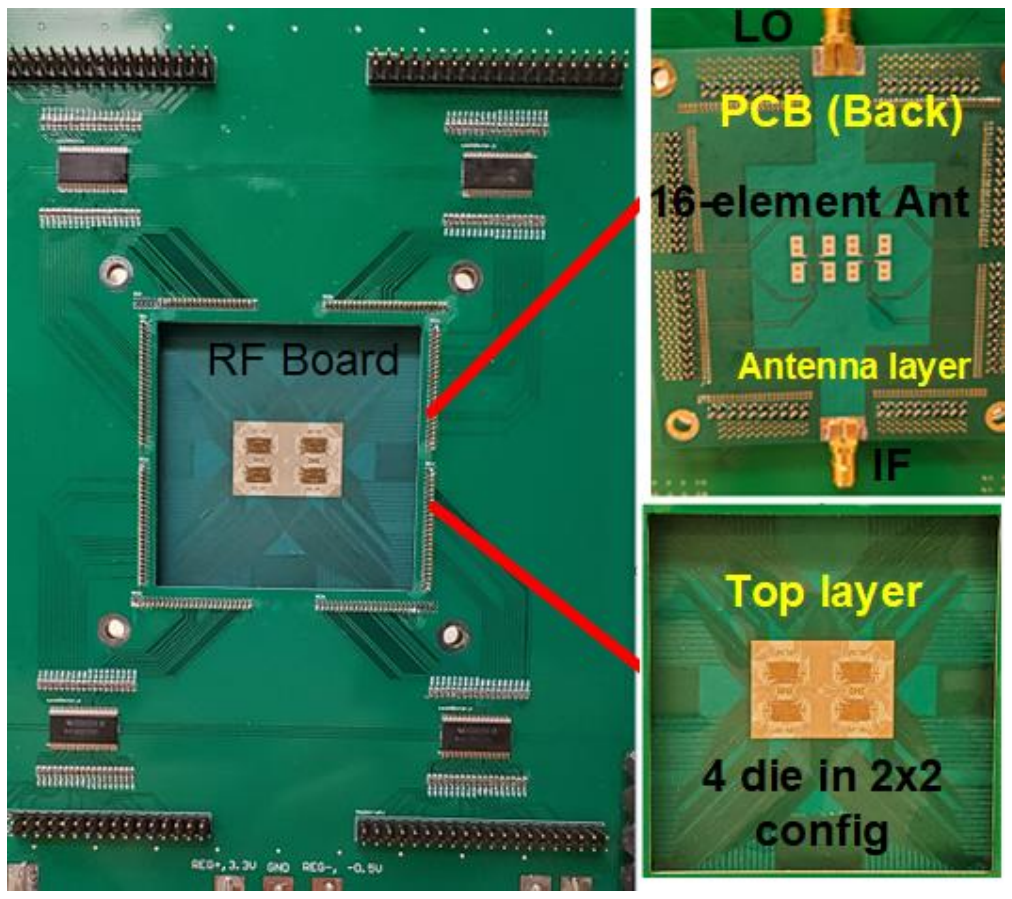

(a)

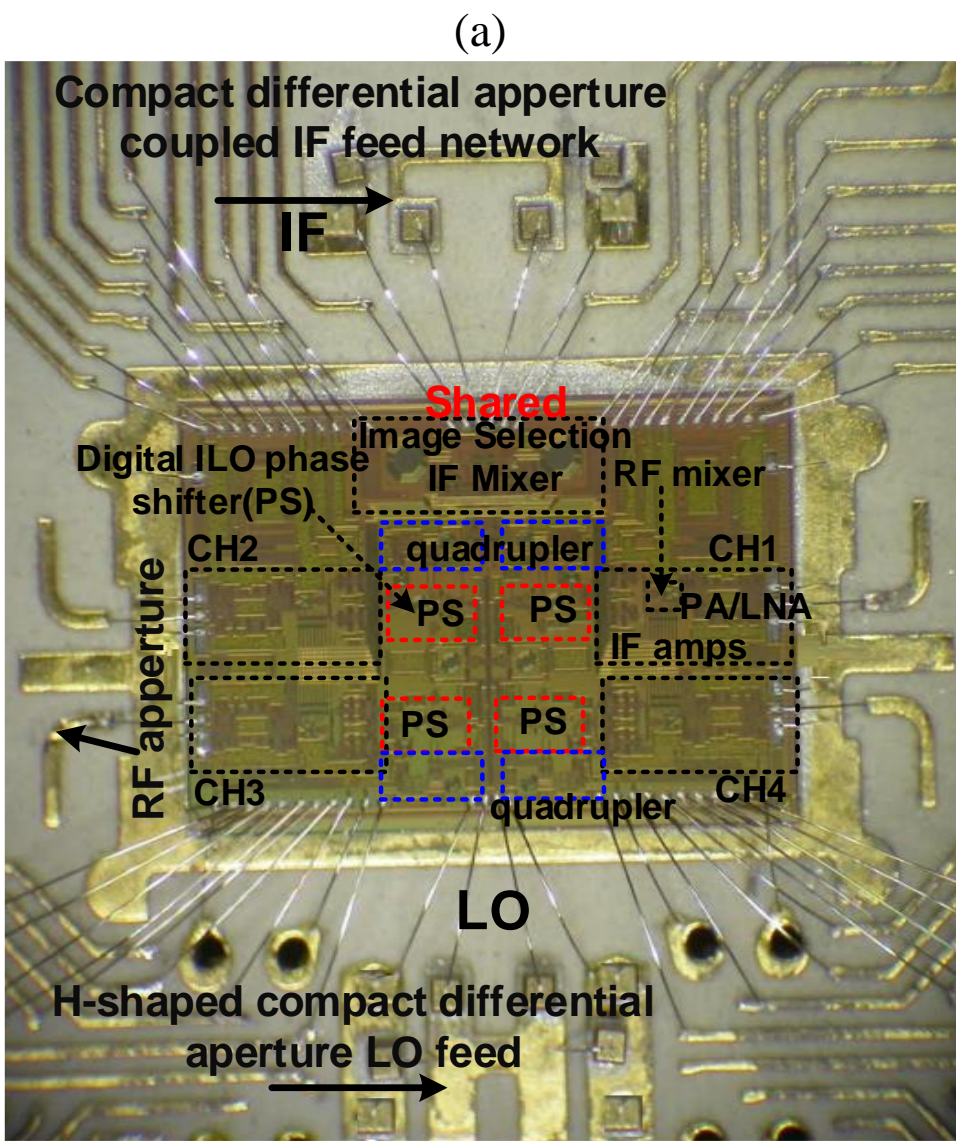

Fig. 10. Fabricated 16-element array, a) the PCB with front and back side view showing 16-element antenna elements and the 4-die placement, respectively, b) micrograph of 2x2 array die along with back and front side views of the implemented 16-element phased array system using four mounted dies.

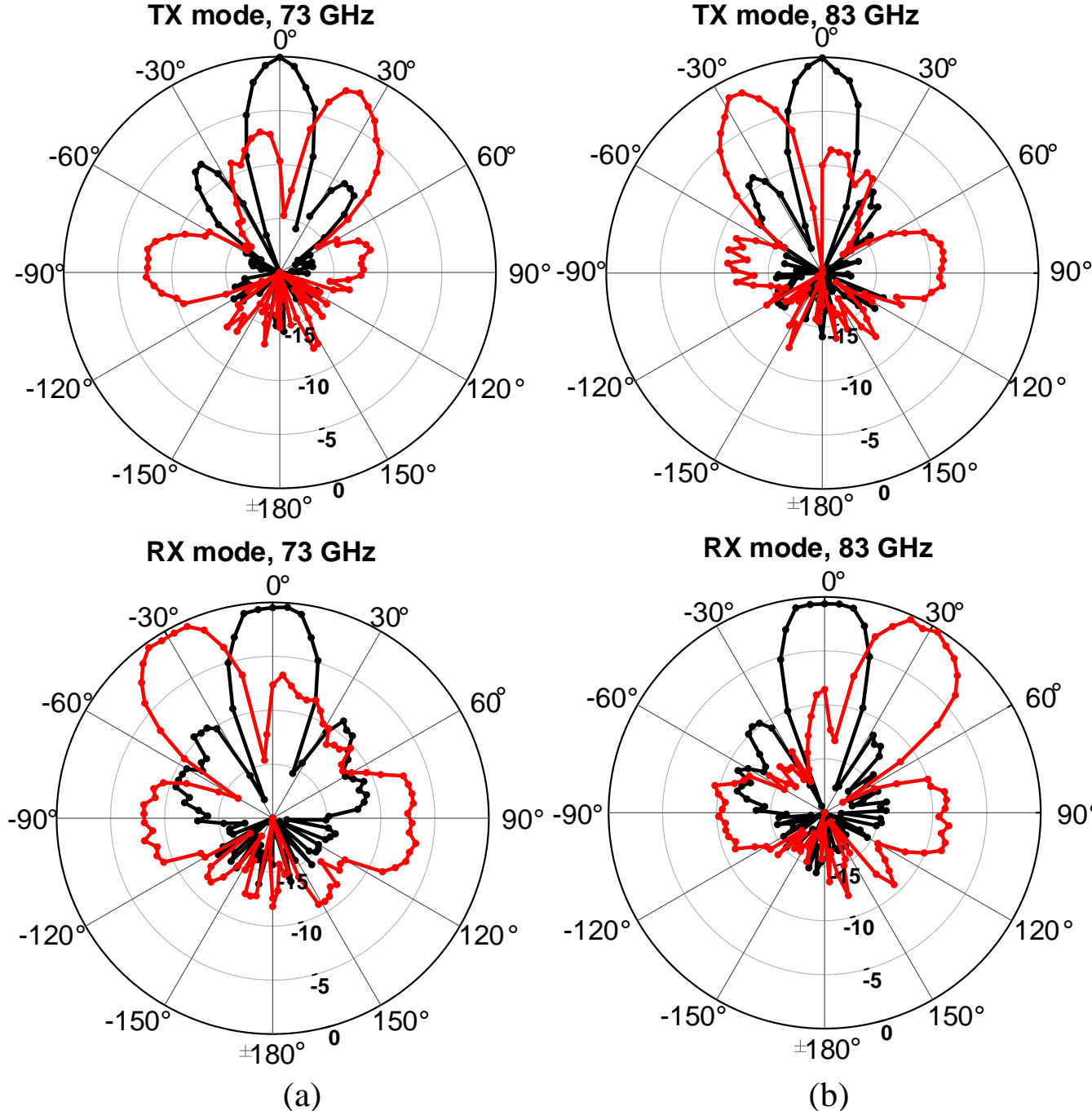

(a) (b)

Fig. 11. Measured beam pattern for 16-element transceiver, TX (black-dot) and RX (red-dot) under two UB and LB modes (a) 73 GHz, (b) 83 GHz.

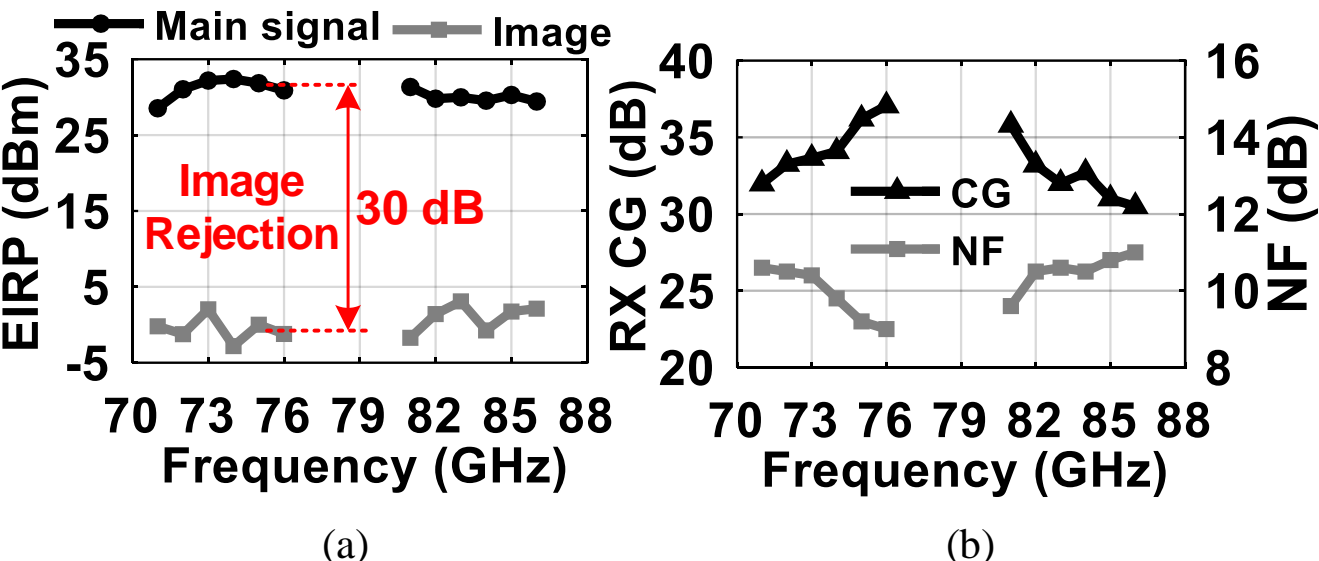

(a) (b)

Fig. 12. (a) Measured 16-element EIRP for main signal (UB/LB) and image signal with one-bit switch of phase inversion for TX mode, (b) measured conversion gain and noise figure for RX mode.

## V. 16-Element Heterogonous Packaging, Implementation and Measurement

The 16-element, 71-86 GHz array, uses 2x2 transceiver die fabricated in 90-nm SiGe BiCMOS [6], has been packaged employing two proposed techniques 1) fixed antenna elements on the antenna layer, 2) differential and miniaturized LO/IF power delivery network, Fig. 10. The chip microphotograph is illustrated in Fig. 10 (b) alongside the 16-element assembled PCB shown in Fig. 10 (a). A single die occupies 4.6 mm x 2.8 mm. The 16-element assembly uses four PCB layers. The top PCB layer routes the RF signals for the 2x2 die while eliminating the need for multi-layer vias with large loss. The aperture layers for IF and LO feed-lines are placed between the RF and LO/IF signals optimized for 5 mil thickness using Rogers RO 3006 substrate. To enhance the radiation gain, a 10-mil Rogers RO3003 substrate with relative permittivity of 3 is used for the antenna and aperture. The resulting aperture coupled antenna array is wideband (10-dB impedance bandwidth), exhibiting a maximum of 2 dB gain variation over 71–86 GHz. The area of the 16-element array is around 11.5 mm×7.1 mm and fits within the $\lambda/2$ spacing limit of the system. The measured antenna gain is 11 to 13 dB due to the low-loss wideband aperture coupled antenna.

The CMOS die illustrated in Fig. 10 (b) is a 4-element, compact bidirectional phased-array transceiver, employing the Weaver image-selection architecture (WISA) that reduces the LO tuning range to 3 GHz (4% FBW) while covering the 10 GHz RF band (20% FBW) [46-47]. It also shares the 1st

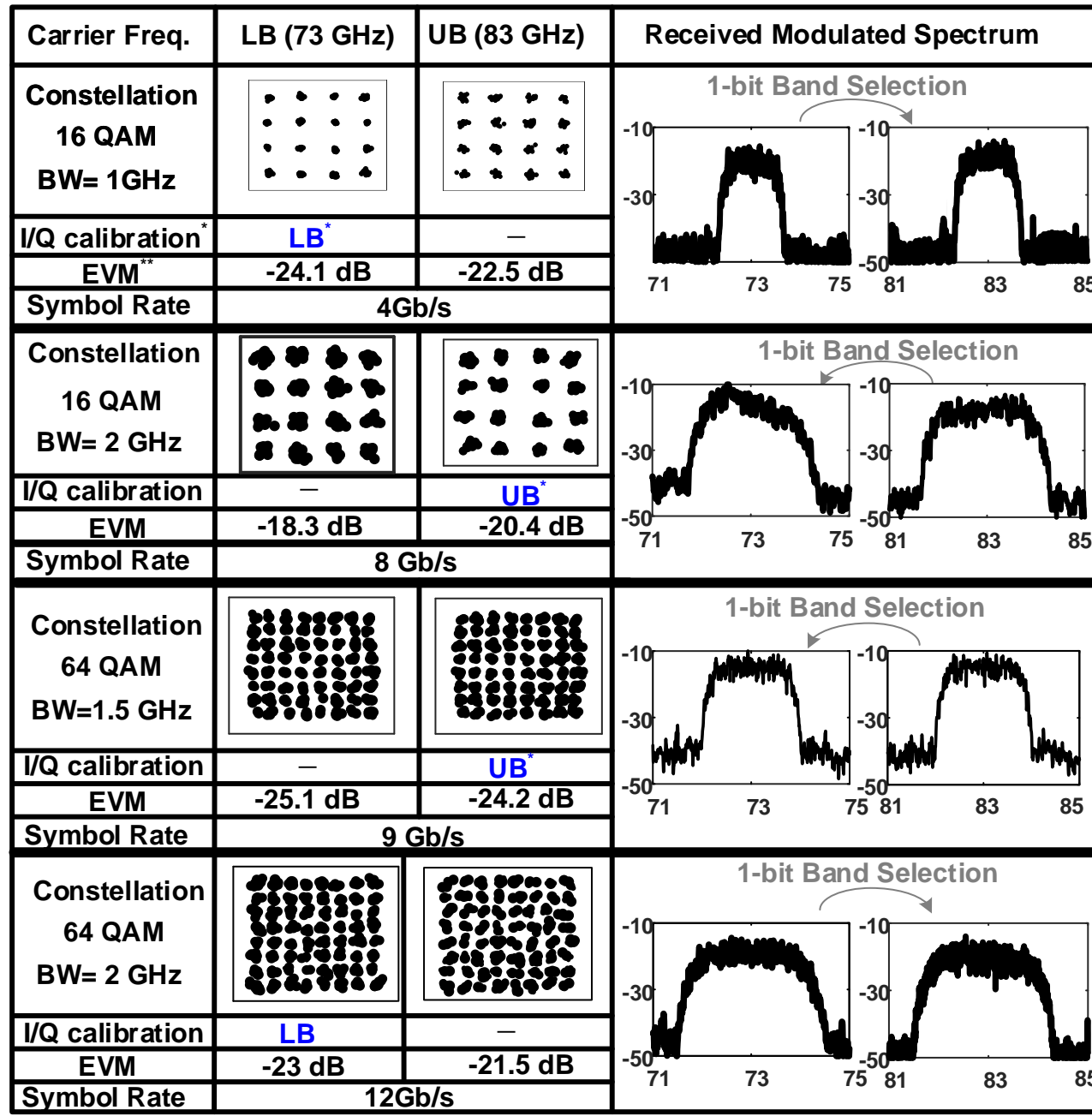

*The calibration is optimized for the one particular band (LB or UB), and the data sent to other band with one-bit phase inverter switch under same calibration state.

(a)

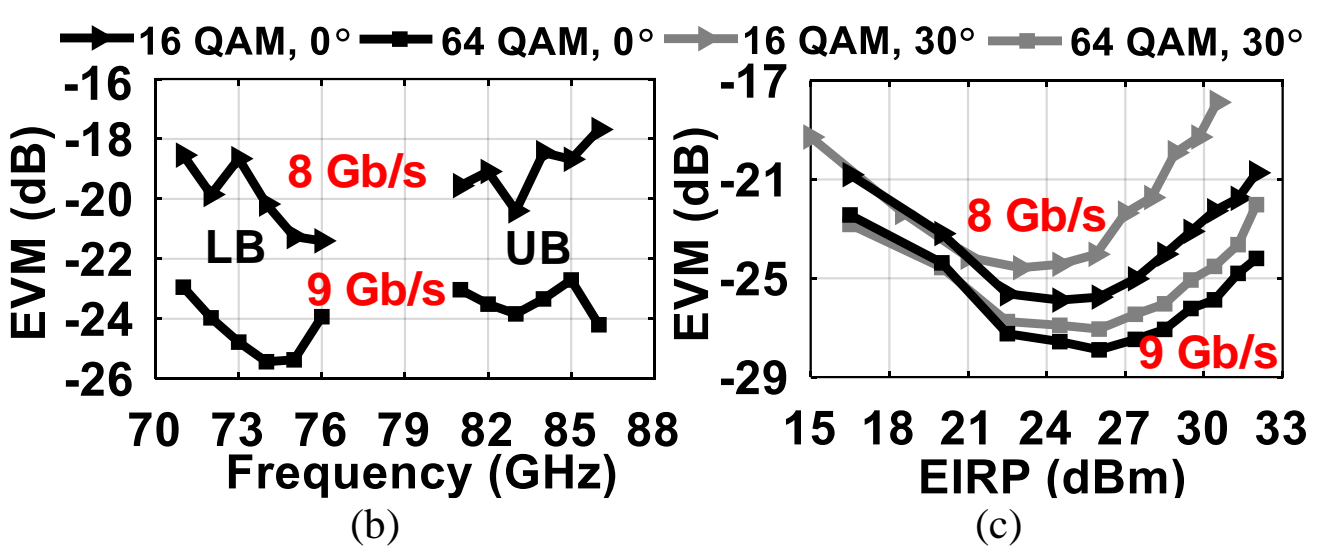

(b) (c)

Fig. 13. (a) The constellation diagram and EVM values for 16 QAM and 64 QAM for two UB and LB under same calibration state, (b) EVM versus frequency for 8Gb/s 16 QAM and 9Gb/s 64 QAM, (c) measured EVM versus output EIRP under 0 and 30 degrees steering angle for 16 QAM and 64 QAM.

intermediate frequency (IF1) between elements to combine the signals into a single bidirectional IF and an image selection stage reduces the number of I/Q mixers. Only one image selection unit with two I/Q filters are required on chip as illustrated in chip layout in Fig. 10 (b) to significantly reduce power consumption and area across the array for TRX modes [6]. By introducing LO phase shifters, the WISA provides beam-steering as a scalable mm-wave array based on LO distribution. The LO is phase-shifted at each element to avoid large RF phase shifters (PS) in each signal path and RF gain variation over the band [48-49]. The LO phase shift is produced at low frequency (19.5 GHz) and frequency-multiplied by four at each element.

The array beam-steering angle for both TX/RX modes and for two lower band (LB) 73.5 GHz, and upper band (UB), 83.5 GHz, are shown in Fig. 11 (a) and (b), respectively, illustrating a maximum ±30° with 15 dB peak to null ratio and 30° half-power beam-width.

The $P_{sat}$ of each element is 8 dBm and the measured EIRP for TX mode is 28 to 32 dBm over the 10 GHz bandwidth, illustrated in Fig. 12 (a). The image signal power is also illustrated in Fig.12 (a) and shows the image rejection with value around 30 dB sufficient for 16 QAM and 64 QAM

Table I. Comparison with State-of-the arts V/E/W bands Phased Array

| | This Work | [16] JSSC | [13] JSSC | [14] ISSCC |
|---|---|---|---|---|
| Technology | 90nm SiGe BiCMOS | 180 nm SiGe | 40 nm CMOS | 22nm FinFET |
| Freq. (GHz) | 71-76<br>81-86 | 80-100 | 57-66 | 71-76 |
| Element per RFIC | 4 | 16TX, 8RX | 12 | 4 |
| Array Size | 16 | 384 | 144 | 64 |
| Architecture | Sliding-IF | Direct I/Q | Direct I/Q | Direct I/Q |
| Beamforming | LO*1 | RF | RF | BB |
| RF BW (GHz)<br>Phase shifter-BW<br>(GHz) (PS/RF (%)) | 10<br>1*2<br>10% | 20<br>20<br>100% | 9<br>9<br>100% | 5<br>2*3<br>40% |
| Antenna | On- PCB | In-package | On- PCB | On- PCB |
| EIRP Total (dBm) | 30 | 60 | 51 | 44.4 |
| TX $P_{sat}$ (dBm) | 8 | 8 | 5 | 8 |
| RX NF (dB) | <9 | 6.5-8 | 7 | 6 |
| RX conv. Gain (dB) | 32 | 80 | 23 | 37 |
| Constellation<br>Data-rate (Gb/s)<br>EVM (dB) | 64 QAM<br>9 Gb/s (-24)<br>16 QAM<br>8 Gb/s (-19) | 64 QAM*4<br>18 Gb/s (-30)<br>16 QAM<br>10 Gb/s (-22) | 16 QAM<br>4.6 Gb/s<br>(-22) | 16 QAM<br>4 Gb/s (-25)<br>16 QAM<br>7.2 Gb/s (-20) |
| TX $P_{DC}$ per el. (mW) | 250 | 275 | 58 | 148 |
| RX $P_{DC}$ per el. (mW) | 160 | 225 | 46 | 168 |
| Variation over RF band (dB) | ± 2 dB | N/A | N/A | - |
| EIRP/$P_{DC}$ x100 | 25% | 9.5% | 15% | 22% |
| Total Die-Area ($mm^2$) | 12.8 | 36.45 | 20 | 5.1 |

1* Digital Injection-locked oscillator phased shifter. *2. The proposed architecture reduces LO FBW to only 4%, resulting in narrow band LO distribution (1GHz). *3: BW of I/Q baseband. *4. According to the referred paper, Shahramian et.al's ISSCC 2018. *Note: the reported EVM of this work in table (-19.5 or -24) are the averaged values over two bands.

waveforms. The RX conversion gain is also plotted in Fig. 12 (b) and is 32 dB on average over the band. The gain variation of RX mode is ±2.5 dB under the constant LO power. However, using variable gain could enhance the gain variation to ±1 dB. The noise figure of the system also measured and varies between 9 dB ($NF_{min}$) to 11 dB over the two bands.

The EVM of the TX and RX is measured and the results for the two specific LB and UB, 73.5 GHz and 83.5 GHz, are illustrated in Fig. 13 (a). The maximum modulation bandwidth achieved for 16 QAM and 64 QAM is 2 GHz (8 Gb/s) and 1.5 GHz (9 Gb/s), respectively . Operation switches from LB to UB with the single bit phase inversion under same calibration state and constant power consumption, resulting in minimum EVM variation. The measured EVM over the two bands with UB/LB selection is presented in Fig. 13 (b) with minimum variation of ± 2 dB across the two bands. An 8-Gb/s, 16-QAM waveform, 2-GHz modulated BW, is transmitted with EVM under -18 dB and an 9-Gb/s 64-QAM waveform is transmitted with EVM under only -23 dB over entire E-band. Due to the narrower modulated BW (1.5 GHz), the 9-Gb/s 64 QAM results in better EVM. The measured EVM over EIRP for 8-Gb/s 16-QAM and 9Gb/s 64-QAM waveforms is also demonstrated in Fig. 13 (c) under beam steering angle of 0 and 30 degrees (max steering angle). The EVM at maximum EIRP and at maximum steering angle, 30 degrees, is degraded by a 2.7 dB and 2 dB for 16 QAM and 64 QAM, respectively.

The proposed bidirectional, wideband and scalable transceiver implemented on miniaturized 4-layer PCB array is compared to prior work in Table *I* and demonstrates the first only four-layer PCB solution to cover both 71-76 and 81-86GHz with 25% EIRP/PDC and minimum ±2dB variation over the bandwidth. The minimum performance variations are mainly originated from the compact and miniaturized innovations from chip level up to package level. The bidirectional shared image selection WISA configuration for the chip circuitry proposed in [6] along with the proposed miniaturized and efficient package layout save the DC power, enhance the EIRP/PDC to 25% and result in ±2dB low EVM variation over the two bands.

## VI. Conclusion

This paper presents novel, miniaturized and efficient architectures for future heterogeneous scalable MMW array. Four dies with a 2x2 configurations are assembled into a 16-element antenna array having a fixed unit-cell antenna for each element with differential excitation of the two half-side of the array. The proposed heterogeneous integration layout requires only four stacking layers and demonstrates novel miniaturized LO and IF distribution between the four dies employing differential aperture coupled feed network. The theoretical analysis shows that aperture coupling technique can be employed for a miniturized and low-error wideband tunable power distributions. The proposed scalable system demonstrates an averaged minimum performance variation of ±2 dB for EVM with 25% EIRP/PDC over the 71-86 GHz band under constant power consumption and same calibration states.

## Acknowledgements

The author thank Dr. James Buckwalter for supporting this work, Dr. Behzad Yektakhah for the help on PCB and Mr. Shah Zaib Aslam for the help with Figure 8 and 9 re-simulations.

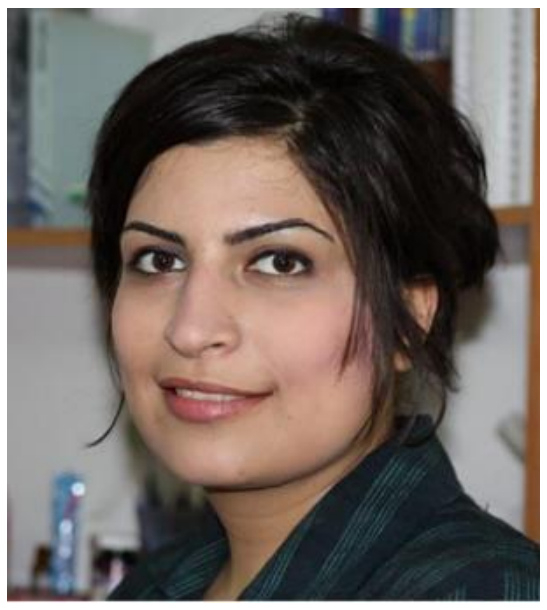

**Najme Ebrahimi** (S'09, M'17) received the B.S. degree (Highest Hons.) in electrical engineering from Shahid Beheshti University, Tehran, Iran, in 2009, the M.S. degree (Highest Hons.) in electrical engineering from the Amirkabir University of Technology, Tehran, in 2011, and the Ph.D. degree in electrical and computer engineering from the University of California at San Diego, La Jolla, CA, USA, in 2017. She was postdoctoral research fellow at the University of Michigan from 2017 to 2020. She is currently an Assistant Professor with the University of Florida, Gainesville, Florida, USA. Her research interests are on RF, millimeter-wave and THz Integrated circuits and systems, communication electronics, wireless communications and sensing, Internet of Things (IoT) connectivity and communications, physical layer security and sensing. Dr. Ebrahimi was a recipient of the Jacobs School of Engineering Fellowship at the University of California at San Diego, 2019 and 2020 EECS RisingStar, 2018-2020 IEEE Microwave Society Chapter Chair for Southeastern Michigan. Dr. Ebrahimi was a recipient of the 2021 Defense Advanced Research Projects Agency (DARPA) Young Faculty Award. She is serving as a technical member of MTT-14 MICROWAVE AND MILLIMETER-WAVE INTEGRATED CIRCUITS.